\documentclass[aps,floats]{revtex4}
\usepackage{amsmath,amssymb}
\usepackage{graphicx,epsfig}

\begin{document}
\bibliographystyle {plain}

\def\oppropto{\mathop{\propto}} 
\def\opsimeq{\mathop{\simeq}}
\def\opoverderline{\mathop{\overline}}
\def\operarroZ{\mathop{\longrightarroZ}}
\def\opsim{\mathop{\sim}} 
\def\opmin{\mathop{\min}} 
\def\opmax{\mathop{\max}} 

\def\fig#1#2{\includegraphics[height=#1]{#2}}
\def\figx#1#2{\includegraphics[width=#1]{#2}}



\title{ Strong Disorder Renewal Approach to DNA denaturation and wetting : \\
typical and large deviation properties of the free energy }


\author{ C\'ecile Monthus }
 \affiliation{Institut de Physique Th\'{e}orique, 
Universit\'e Paris Saclay, CNRS, CEA,
91191 Gif-sur-Yvette, France}
 
\begin{abstract}
For the DNA denaturation transition in the presence of random  contact energies, or equivalently the disordered wetting transition, we introduce a Strong Disorder Renewal Approach to construct the optimal contacts in each disordered sample of size $L$. The transition is found to be of infinite order, with a correlation length diverging with the essential singularity $\ln \xi(T) \propto \vert T-T_c \vert^{-1}$. In the critical region, we analyze the statistics over samples of the free-energy density $f_L$ and of the contact density, which is the order parameter of the transition. At the critical point, both decay as a power-law of the length $L$ but remain distributed, in agreement with the general phenomenon of lack of self-averaging at random critical points. We also obtain that for any real $q>0$, the moment $\overline{Z_L^q} $ of order $q$ of the partition function at the critical point is dominated by some exponentially rare samples displaying a finite free-energy density, i.e. by the large deviation sector of the probability distribution of the free-energy density.

\end{abstract}

\maketitle

\section{Introduction } 

Wetting transitions are in some sense the simplest classical phase transitions,
since they involve linear systems \cite{mfisher}. 
The effects of quenched disorder on the wetting transition 
or on the equivalent Poland-Scheraga model of DNA denaturation \cite{poland,Pol_Scher}
 have thus attracted a lot of interest among physicists
 \cite{luck,hakim,hwa,tang,kafri,azbel,barbara,thomas,kunz,retaux}
and mathematicians (see the books \cite{giacomin} and references therein).
Within the field of disordered models,
the DNA denaturation model is very special, 
because the system can avoid some disorder variables by making loops,
whereas spin models have to cope with all the random couplings whatever they are.
In particular, in the critical region where the contact density vanishes,
each configuration involves only a vanishing fraction of disorder variables.
As a consequence, Tang and Chat\'e \cite{tang} have proposed
that the denaturation transition is driven by the rare anomalously attractive regions :

(i) on one hand, they have proposed a simple scaling argument for a system of size $L$, based on the competition
between the energy gain of the best attractive segment and the entropic cost of a system-size loop :
both scale as the logarithm $\ln L$ of the size $L$, so that
this argument points towards an essential singularity divergence of the correlation length
\begin{eqnarray}
\xi(T) \propto e^{\frac{B}{\vert T-T_c \vert}} 
\label{xitessential}
\end{eqnarray}
This behavior is reminiscent of the critical properties found in the different model
concerning a polymer at the interface between two selective solvents \cite{huse,c_rgcopolymer} for similar reasons.
 It turns out that the singularity of Eq. \ref{xitessential} has also been found recently
in the quantum phase transition of random transverse field Ising model with
long-ranged power-law couplings \cite{igloi_lr,igloi_lr3d} (or in long-ranged epidemic models in a random environment \cite{igloi_epidemic})
and can be also explained in terms of Extreme Value Statistics  \cite{igloi_lr,igloi_lr3d}.

(ii) on the other hand, Tang and Chat\'e \cite{tang} and more recently Derrida and Retaux \cite{retaux} 
have concluded that the real space renormalization on hierarchical lattices 
 leads instead to the Berezinski-Kosterlitz-Thouless (B.K.T.) essential singularity
\begin{eqnarray}
\xi_{BKT}(T) \propto e^{\frac{B}{\sqrt{\vert T-T_c \vert}}} 
\label{xiBKT}
\end{eqnarray}
This B.K.T. scaling emerges from the joint renormalization flows of an amplitude
and of an exponential rate \cite{tang,retaux}.
We are not aware of a simple scaling physical argument that would explain the origin of the additional square root in Eq. \ref{xiBKT}
with respect to Eq. \ref{xitessential}.

In this paper, we introduce a Strong Disorder Renewal Approach that 
can be considered as an elaboration of the scaling argument (i) :
it leads to the essential singularity of Eq. \ref{xitessential},
but in addition, it allows to compute explicitly many observables, including
the statistics of the free-energy and of the contact density over the disordered samples
of a given size $L$.
 The paper is organized as follows.
After the description of the model in section \ref{sec_model},
we explain the Strong Disorder Renewal Approach in section \ref{sec_renewal}.
We then analyze the statistics of the loop lengths in section \ref{sec_loops},
the statistics of the contact density in section \ref{sec_contacts},
and the statistics of the free-energy in section \ref{sec_free} :
typical and large deviation properties are given respectively
for the delocalized phase in section \ref{sec_freedeloc},
for the critical point in section \ref{sec_freecriti},
and for the localized phase in section \ref{sec_freeloc}.
The validity of the Strong Disorder Renewal Approach is discussed in section \ref{sec_validity}.
Our conclusions are summarized in section \ref{sec_conclusion}.
The Appendix \ref{sec_app} explains how the Strong Disorder Renewal Approach
can be adapted to other distributions of the contact energies, while the main text focuses on the 
simplest case of the exponential distribution.

\section{ Model and notations} 

\label{sec_model}

\subsection{ Partition function } 

We consider a polymer of length $L$ attached at the origin $n_0=0$ and free at the other end :
the partition function is a sum over the number $K=0,1,..,L$ of contacts and of their positions $0<n_1<n_2<..<n_K \leq L $
\begin{eqnarray}
Z(L) =  \sum_{K=0}^{L}  
 \sum_{n_0=0<n_1<n_2<..<n_K \leq L} Z_K(n_1,n_2,..,n_k) 
\label{zfull}
\end{eqnarray}
of
\begin{eqnarray}
 Z_K(n_1,n_2,..,n_k) =\Omega(n_1 )e^{ \frac{\epsilon_{n_1}}{T} }
\Omega(n_2-n_{1} )e^{ \frac{\epsilon_{n_2}}{T} } ...
 \Omega(n_K-n_{K-1} )e^{ \frac{\epsilon_{n_K}}{T} } 
\label{zK}
\end{eqnarray}
The contact energies $\epsilon_n$ are random quenched variables, while 
the weight $\Omega(l)  $ of a loop of length $l$ between two contacts displays the power-law behavior for large $l$
(here to simplify the notations, we will consider that this power-law holds for all lengths $l=1,2,..$)
\begin{eqnarray}
\Omega(l) = \frac{1}{l^c} 
\label{boucle}
\end{eqnarray}
The exponent $c$ is obtained from the enumeration of random walks returning to the origin
and thus depends on the assumptions made concerning self-avoidance and excluded-volume interactions \cite{peliti}.
Here we wish to consider $c$ as a free parameter, that can take large values (for reasons that will be discussed in section \ref{sec_validity}),
so we may for instance consider usual random walks in dimensions $d \geq 2$ \cite{peliti}
\begin{eqnarray}
c^{RW} = \frac{d}{2}
\label{cRW}
\end{eqnarray}

From the form of the partition function, it is thus clear that the DNA denaturation transition or the wetting transition 
corresponds to a competition between the contact energies that are 'good' $\epsilon_n>0 $
and the entropic logarithmic cost $ \ln \Omega(l) = -c \ln l <0 $ of loops between contacts. 

\subsection{ Probability distribution of the random contact energies } 

 In the main text of the present paper, we focus on the case where the contact energies $\epsilon_{n} $ are independent random variables,
drawn with the exponential distribution of parameter $W$ 
(see the Appendix \ref{sec_app} for the case of other distributions)
\begin{eqnarray}
\rho(\epsilon) = \frac{1}{W} e^{-  \frac{(\epsilon-\epsilon_{min})}{W} } \theta(\epsilon \geq \epsilon_{min})
\label{rhoexp}
\end{eqnarray}
with some lower value $\epsilon_{min}<0$ to ensure the presence of repulsive contacts $\epsilon<0$. The proportion of attractive contacts $\epsilon>0$
\begin{eqnarray}
A \equiv \int_0^{+\infty} d \epsilon \rho(\epsilon) =  e^{ \frac{\epsilon_{min}}{W} } 
\label{amplitude}
\end{eqnarray}
can be chosen anywhere in the interval
\begin{eqnarray}
0<A<1
\label{amplitude01}
\end{eqnarray}
In the following, an essential role will be played by the proportion of contact energies above an arbitrary threshold $\eta>0$
\begin{eqnarray}
\int_{\eta}^{+\infty} d \epsilon \rho(\epsilon) = A e^{-  \frac{\eta}{W} } 
\label{rhotail}
\end{eqnarray}

\subsection{ Phase transition criterion based on the free-energy  }

Instead of the true free-energy $F^{true}(L)$ of each sample, it has become usual in this field to introduce the more convenient notation
\begin{eqnarray}
F(L) \equiv \ln Z(L) = -\beta F^{true}(L)
\label{freeenergy}
\end{eqnarray}
as well as its density per unit length
\begin{eqnarray}
f(L) \equiv \frac{F(L)}{L} = \frac{\ln Z(L)}{L}
\label{freeenergydensity}
\end{eqnarray}

Then the localized phase corresponds to a positive limit 
\begin{eqnarray}
 f^{loc}(L \to +\infty)  > 0
\label{freedensityloc}
\end{eqnarray}
while the delocalized phase corresponds to a vanishing limit 
\begin{eqnarray}
 f^{deloc}(L\to +\infty) = 0
\label{freedensitydeloc}
\end{eqnarray}

Besides this thermodynamic limit, 
we will be interested into the probability distribution over samples
of the free-energy in finite size $L$, in particular at the critical point.

\subsection{ Order parameter of the phase transition  } 

In a given sample, the averaged number $<K_L> $ of contacts is computed as
\begin{eqnarray}
<K_L> \equiv  \frac{ \sum_{K=0}^{L}  K \sum_{n_0=0<n_1<n_2<..<n_K<L} Z_K(n_1,n_2,..,n_k) }{\sum_{K=0}^{L}  \sum_{n_0=0<n_1<n_2<..<n_K<L} Z_K(n_1,n_2,..,n_k)} 
\label{contactKL}
\end{eqnarray}

The order parameter of the transition is the contact density, i.e. the number of contacts per unit length
\begin{eqnarray}
k(L) \equiv \frac{<K_L>}{L} 
\label{contactdensity}
\end{eqnarray}
that remains finite in the localized phase as $L \to +\infty$
\begin{eqnarray}
  k^{loc}(L\to +\infty) >0
\label{contactdensityloc}
\end{eqnarray}
while it vanishes in the delocalized phase
\begin{eqnarray}
 k^{deloc}(L\to +\infty)  =0
\label{contactdensitydeloc}
\end{eqnarray}
Again, besides this thermodynamic limit,
 it is interesting to analyze the probability distribution over samples
of the number of contacts in finite size $L$, in particular at the critical point.

\section{ Strong Disorder Renewal Approach   } 

\label{sec_renewal}

Strong Disorder Approaches are based on the general idea that
the spatial heterogeneities introduced by the quenched disorder variables
dominate over quantum, thermal, or stochastic fluctuations depending on the
considered model (see for instance the review \cite{review_igloi}).
While these Strong Disorder Approaches are usually formulated within
real-space renormalization procedures \cite{review_igloi},
as for the polymer at the interface between two selective solvents \cite{c_rgcopolymer},
we propose in this section a Strong Disorder Renewal Approach
in each disordered sample.

\subsection{   Strategy in each disordered sample } 

\label{sec_strategy}

In a disordered sample corresponding to 
a given realization  $(\epsilon_1,...,\epsilon_L)$
of the random contact energies, we consider the following strategy :

(i) the first return $n_1^*$ takes place at the first 'good enough contact',
defined as the first position where $\Omega(n) e^{ \frac{\epsilon_n} {T}}>1$,
 i.e. where the contact energy is bigger than the entropic cost of the return
\begin{eqnarray}
\epsilon_{n_1^*} > - T \ln \Omega (n_1^*)
\label{first}
\end{eqnarray}
while all the previous $(n_1^*-1)$ positions $n=1,..,n_1^*-1$ satisfy
\begin{eqnarray}
\epsilon_{n} < - T \ln \Omega (n)
\label{firsto}
\end{eqnarray}

The corresponding gain for the logarithm of the partition function is
\begin{eqnarray}
f_{n_1^*}= \frac{\epsilon_{n_1^*}}{T} + \ln \Omega (n_1^*) \geq 0
\label{x1gain}
\end{eqnarray}

(ii) once the first  'good enough contact' has been found at $n_1^*$, one may iterate the same procedure:
 the second return $n_2^*$ takes place at the next 'good enough contact',
define as the position $n_2^*$ satisfying
\begin{eqnarray}
\epsilon_{n_2^*} > - T \ln \Omega (n_2^*-n_1^*)
\label{second}
\end{eqnarray}
while all the intermediate $(n_2^*-n_1^*-1)$ positions $n=n_1^*+1,..,n_2^*-1$ satisfy
\begin{eqnarray}
\epsilon_{n} < - T \ln \Omega (n-n_1^*)
\label{secondo}
\end{eqnarray}

The corresponding gain for the logarithm of the partition function is
\begin{eqnarray}
f_{n_2^*}= \frac{\epsilon_{n_2^*}}{T} + \ln \Omega (n_2^*-n_1^*) \geq 0
\label{x2gain}
\end{eqnarray}

This strategy thus define a very simple renewal process.

\subsection{ Partition function and order parameter in each sample  } 

So in each given sample of length $L$, one ends up with a certain number $K^*$
of 'good enough contacts' located at the positions $(n_1^*,..,n_K^*)$, and
one considers that the partition function of this sample (Eq \ref{zfull})
 is completely dominated by this optimal configuration
\begin{eqnarray}
 Z(L) \simeq    Z_{K^*}(n_1^*,n_2^*,..,n_k^*) = e^{\sum_{k=1}^{K^*} f_{n_k^*} }
\label{zopt}
\end{eqnarray}
Its logarithm (Eq. \ref{freeenergy}) corresponds to the sum of the gains $f_{n_k^*}$ of the $K^*$ contacts
\begin{eqnarray}
F(L) \equiv \ln Z(L) \simeq \sum_{k=1}^{K^*} f_{n_k^*} 
\label{freeopt}
\end{eqnarray}
The thermally averaged number of contacts of the sample (Eq. \ref{contactKL}) is simply
\begin{eqnarray}
<K_L> \simeq K^*
\label{contactopt}
\end{eqnarray}

In this paper, our goal is to analyze how this number $K^*$ of contacts and the corresponding free-energy of Eq. \ref{freeopt}
are distributed over the disordered samples of a given length $L$.

\subsection{ Statistics of the number $K^*$ of contacts over the disordered samples of length $L$  }

The probability distribution $\pi_{K^*}(L)$ of the number $K^*$ of contacts 
over the disordered samples of length $L$
is normalized to
\begin{eqnarray}
\sum_{K^*=0}^L \pi_{K^*}(L) =1
\label{normapik}
\end{eqnarray}

A related important observable is the probability distribution $P(l)$
of the loop length $l=1,2,..$ between two contacts: 
 if one adds the initial value $\pi_0(0)=1$, 
the probability $\pi_{K^*=0}(l)$ of zero contact plays the role of the cumulative
distribution for $P(l)$ that can be computed as the difference
\begin{eqnarray}
P(l)= \pi_0(l-1)-\pi_0(l)
\label{pgood}
\end{eqnarray}
and one has the sum rule
\begin{eqnarray}
\sum_{l=1}^L P(l)= 1-\pi_0(L)
\label{sumrule}
\end{eqnarray}
In the limit $L \to +\infty$, the normalization thus reads
\begin{eqnarray}
\sum_{l=1}^{+\infty} P(l) = 1- \pi_0(\infty)
\label{normapl}
\end{eqnarray}
where $\pi_0(\infty)$ represents the probability to find zero good contact for $L \to +\infty$ : 
it vanishes in the localized phase 
\begin{eqnarray}
\pi_0^{loc}(L \to \infty)=0
\label{pi0linftyloc}
\end{eqnarray}
but remains finite in the delocalized phase
\begin{eqnarray}
 \pi_0^{deloc}(L \to \infty) >0
\label{pi0linftydeloc}
\end{eqnarray}

The probability of $K^*=1$ contact can be then computed as the convolution
\begin{eqnarray}
\pi_1(L) = \sum_{l=1}^L P(l) \pi_0(L-l)
\label{pi1}
\end{eqnarray}
Similarly for an arbitrary number of contacts $1 \leq K^* \leq L$, the probability reads
\begin{eqnarray}
\pi_{K^*}(L) && = \sum_{l_1\geq 1} \sum_{l_2 \geq 1} ... \sum_{l_{K^*} \geq 1} \sum_{l \geq 0} P(l_1) P(l_2) ..  P(l_{K^*}) \pi_0(l) 
\delta \left(L-l -\sum_{k=1}^{K^*} l_k\right)
\label{pik}
\end{eqnarray}
and will be thus closely related to the statistics of the sum of loop lengths.

\subsection{ Statistics of the partition function over the disordered samples
of length $L$  } 

For a given sample of length $L$ displaying $K^*$ contacts, 
the logarithm of the partition function is the sum over the $K^*$ independent gains $f_{n_k^*}$ of the contacts (Eq. \ref{freeopt}).
If one introduces the probability distribution $p_K(F)$ of the sum of $K$ independent variables $f$ distributed with $p_1(F)$,
the probability distribution ${\cal F}_L(F)$ of $F=\ln Z$ over the samples of length $L$ 
can be written using the probability $\pi_{K^*}(L)$ of the number $K^*$ of contacts introduced above
\begin{eqnarray}
{\cal F}_L(F) =  \sum_{K^*=0}^{L} \pi_{K^*}(L)  p_{K^*}(F)
\label{freeproba}
\end{eqnarray}

Besides the typical region for the free-energy, 
it will be also interesting to discuss
 the behavior of the moments of the partition function for any real value $q>0$
\begin{eqnarray}
M_q(L) \equiv \overline{Z^q(L) } = \overline{e^{q F(L) } }  = \sum_{K^*=0}^{L} \pi_{K^*}(L) \left[  \overline{e^{q x } }     \right]^{K^*}
= \sum_{K^*=0}^{L} \pi_{K^*}(L) \left[ \int_0^{+\infty} df e^{qf} p_1(f)     \right]^{K^*}
\label{momentq}
\end{eqnarray}
that will depend on the large-deviation properties of the probability distribution of the free-energy.

\section{ Statistics of the loop lengths  }

\label{sec_loops}

In this section, we discuss the probability distribution $P(l)$ of the loop length $l$ (Eq. \ref{pgood})
and its cumulative distribution $\pi_0(l)$ (Eq \ref{sumrule}).

\subsection{ Probability $\pi_0(L)$ of zero contact in $L$ }

The probability $\pi_0(L)$ of $K^*=0$ contact in $L$ is the probability that
all energies $\epsilon_n$ for $n=1,2,..,L$ satisfy Eq. \ref{firsto}
\begin{eqnarray}
\pi_0(L) = \prod_{n=1}^L \left[ 1- \int_{-T \ln \Omega (n) }^{+\infty} d \epsilon\rho(\epsilon) \right]
\label{cumulative}
\end{eqnarray}
It is thus convenient to rewrite its logarithm as a sum
\begin{eqnarray}
\ln \pi_0(L) && =- \sum_{n=1}^L u(n)
\label{cumulativelog}
\end{eqnarray}
where the elementary term $u(n)$ reads for the model that we consider (Eqs \ref{boucle} and \ref{rhotail})
\begin{eqnarray}
u(n) \equiv - \ln \left[ 1- \int_{-T \ln \Omega (n) }^{+\infty} d \epsilon\rho(\epsilon) \right]
= - \ln \left[ 1- \frac{ A }{n^{ \frac{c T }{W} } } \right]
\label{un}
\end{eqnarray}

For large $n$, it decays as the power-law
\begin{eqnarray}
u(n) \opsimeq_{n \to +\infty}  \frac{ A }{n^{ \frac{c T }{W} } } = \frac{ A }{n^{ \frac{ T }{T_c} } }
\label{ularge}
\end{eqnarray}
where we have introduced the critical temperature
\begin{eqnarray}
T_c = \frac{W}{c}
\label{tc}
\end{eqnarray}
that will separate the regions of convergence and divergence of the series of Eq. \ref{cumulativelog},
since $\pi_0(L \to +\infty)$ is the simplest criterion of the transition (Eq \ref{pi0linftyloc} and \ref{pi0linftydeloc})
within the present framework.

\subsection{ Loop length statistics at the critical point $T=T_c$  }

For $T=T_c$, the asymptotic behavior of Eq \ref{ularge}
\begin{eqnarray}
u^{criti}(n) \opsimeq_{n \to +\infty}   \frac{A}{n }
\label{ulargecriti}
\end{eqnarray}
yields the logarithmic divergence of the series of Eq. \ref{cumulativelog}
\begin{eqnarray}
\ln \pi_0^{criti}(L) = - \sum_{n=1}^L u^{criti}(n) \opsimeq_{L \to +\infty}  -  A \ln L
\label{cumulativelogcriti}
\end{eqnarray}
corresponding to the power-law decay
\begin{eqnarray}
 \pi_0^{criti}(L) \oppropto_{L \to +\infty}  \frac{1}{L^{A} }
\label{cumulativecriti}
\end{eqnarray}

The probability distribution of the loop length (Eq. \ref{pgood}) decays as
\begin{eqnarray}
P^{criti}(l)= \pi_0^{criti}(l-1)-\pi_0^{criti}(l)  \oppropto_{l \to +\infty}   \frac{A}{l^{1+A} } 
\label{pgoogcriti}
\end{eqnarray}
Since $0<A<1$ (Eq. \ref{amplitude01}), the averaged length diverges
\begin{eqnarray}
l^{criti}_{av} \equiv \sum_{l=1}^{+\infty} l P^{criti}(l)= \infty
\label{lavcriti}
\end{eqnarray}

\subsection{ Localized phase $T<T_c$  }

In the localized phase $T<T_c$, it is convenient to use the reduced temperature
\begin{eqnarray}
t \equiv 1- \frac{T}{T_c} 
\label{tloc}
\end{eqnarray}
that belongs to the interval $0<t<1$.

The asymptotic behavior of Eq \ref{ularge}
\begin{eqnarray}
u^{loc}(n) \opsimeq_{l \to +\infty}   \frac{A  }{n^{\frac{ T }{T_c}} } = \frac{A}{n^{1-t} }
\label{ulargeloc}
\end{eqnarray}
yields the power-law divergence of the series of Eq. \ref{cumulativelog}
\begin{eqnarray}
\ln \pi_0^{loc}(L) = - \sum_{n=1}^L u^{loc}(n) \opsimeq_{L \to +\infty}  -  \frac{A L^t }{ t }
\label{cumulativelogloc}
\end{eqnarray}
corresponding to the stretched exponential decay of exponent $t$
\begin{eqnarray}
 \pi_0^{loc}(L) \oppropto_{L \to +\infty}  e^{ -  \frac{A L^t }{ t } }
\label{cumulativeloc}
\end{eqnarray}

The probability distribution of the loop length (Eq. \ref{pgood}) decays with the same stretched exponential 
\begin{eqnarray}
P^{loc}(l)= \pi_0^{loc}(l-1)-\pi_0^{loc}(l)  \opsimeq_{l \to +\infty}   \frac{A}{l^{1-t} } e^{-  \frac{A l^t }{ t } }
\label{pgoodloc}
\end{eqnarray}
so that all moments are finite.
It is interesting to compute how these moments of arbitrary order $p$ diverge as $t $ becomes small
as a consequence of the decay of Eq. \ref{pgoodloc}
\begin{eqnarray}
l_{p}^{loc} && \equiv \sum_{l=1}^{+\infty} l^p P^{loc}(l) 
 \simeq A  \int_1^{+\infty} dl \  l^{p+t-1}     e^{-  A \frac{l^t-1 }{ t } } 
\nonumber \\
&&= \frac{1}{ t } \int_0^{+\infty} dv \left( 1+\frac{v}{A} \right)^{\frac{p}{t}} e^{- \frac{v}{t} } 
= \frac{1}{ t } \int_0^{+\infty} dv e^{ \frac{h(v)}{t} }
\label{lpcalculloc}
\end{eqnarray}
where the saddle function $h(v)$ and its two first derivatives read
\begin{eqnarray}
h(v) && = p\ln \left( 1+\frac{v}{A} \right) -v 
\nonumber \\
h'(v) && = \frac{p}{A+v}-1 
\nonumber \\
h''(v) && = -\frac{p}{(A+v)^2} 
\label{phiv}
\end{eqnarray}
The saddle-point $v^*$ of the integral of Eq. \ref{lpcalculloc} corresponding to $ \phi'(v^*)$ is
\begin{eqnarray}
v^*=p-A
\label{vsaddle}
\end{eqnarray}
For $p>A$, the saddle-point value belongs to the domain of integration.
Using the values
\begin{eqnarray}
h(v^*) && = p \ln \left( 1+\frac{v^*}{A} \right) -v^* = p \ln \frac{p}{A} -p+A
\nonumber \\
h''(v^*) && = -\frac{p}{(A+v^*)^2} =-\frac{1}{p} 
\label{phivstar}
\end{eqnarray}
the saddle-point evaluation leads to the essential singularity divergences
\begin{eqnarray}
l_{p}^{loc} 
&& \opsimeq_{t \to 0}  \sqrt{ \frac{2 \pi p}{t}}  e^{ \frac{p}{t} \left[ \frac{A}{p}-1-\ln \frac{A}{p}  \right] }
\label{nploc}
\end{eqnarray}
In particular, the averaged length corresponding to the special case $p=1$ diverges as
\begin{eqnarray}
l_{av}^{loc} \equiv \sum_{l=1}^{+\infty} l P^{loc}(l) =  l_{p=1}^{loc}
&& \opsimeq_{t \to 0}  \sqrt{ \frac{2 \pi }{t}}  e^{ \frac{1}{t} \left[ A-1-\ln A  \right] }
\label{navloc}
\end{eqnarray}

\subsection{ Delocalized phase $T>T_c$  }

In the delocalized phase $T>T_c$, it is convenient to use the reduced temperature
\begin{eqnarray}
\theta \equiv \frac{T}{T_c}-1 >0
\label{tdeloc}
\end{eqnarray}

The power-law decay of Eq \ref{ularge}
\begin{eqnarray}
u^{deloc}(l) \opsimeq_{l \to +\infty}   \frac{A}{l^{\frac{ T }{T_c}} } = \frac{A}{l^{1+\theta} }
\label{ulargedeloc}
\end{eqnarray}
yields the convergence of the series of Eq. \ref{cumulativelog} as
\begin{eqnarray}
\ln \pi_0^{deloc}(L) \opsimeq_{L \to +\infty} - A \frac{1-L^{-\theta}}{\theta} 
\label{cumulativelogdeloc}
\end{eqnarray}
The finite limit as $L \to +\infty$ displays the essential singularity as $\theta \to 0$
\begin{eqnarray}
 \pi_0^{deloc}(L=\infty) \opsimeq_{\theta \to 0} e^{- \frac{A}{\theta}}
\label{cumulativelogdelocfinite}
\end{eqnarray}

The probability distribution of the loop length (Eq. \ref{pgood}) decays as
\begin{eqnarray}
P^{deloc}(l)= \pi_0^{deloc}(l-1)-\pi_0^{deloc}(l)  \opsimeq_{l \to +\infty}
   \frac{A}{l^{1+\theta} }  e^{- \frac{A}{\theta}}
\label{pgooddeloc}
\end{eqnarray}
Near the critical point where $0<\theta<1$, the averaged length diverges
\begin{eqnarray}
l^{deloc}_{av} \equiv \sum_{l=1}^{+\infty} l P^{deloc}(l)= \infty
\label{lavdeloc}
\end{eqnarray}

\subsection{  Finite-size scaling in the critical region  }

In the critical region, the above results concerning the probability $\pi_0(L)$
of $K=0$ contacts can be summarized into
\begin{eqnarray}
\ln \pi_0(L)
&&  \simeq - \sum_{l=1}^L A l^{- \frac{T}{T_c}} \simeq  - A  \frac{ L^{1-\frac{T}{T_c}} -1 }{ 1-\frac{T}{T_c} } 
\label{pi0criti}
\end{eqnarray}
This corresponds to the following finite-size scaling form involving the logarithm $(\ln L)$ of the system-size $L$ (instead of 
the usual power-law $L^{\frac{1}{\nu_{FS}}}$ with some finite size correlation length exponent $\nu_{FS}$)
\begin{eqnarray}
\ln \pi_0(L) \opsimeq_{L \to +\infty} 
&&   - A \ln L \ \ \psi \left[ v \equiv \left( 1-\frac{T}{T_c} \right) \ln L \right]
\label{fss}
\end{eqnarray}
where the scaling function 
\begin{eqnarray}
\psi(v) =    \frac{ e^{v} -1 } {v } 
\label{fsspsi}
\end{eqnarray}
is unity at criticality 
\begin{eqnarray}
\psi(v=0) =1
\label{fsspsicriti}
\end{eqnarray}
 exponentially large in the localized phase
\begin{eqnarray}
\psi(v) \opsimeq_{v \to +\infty}    \frac{ e^{v}  } {v } 
\label{fsspsiloc}
\end{eqnarray}
and decays as a power-law in the delocalized phase
\begin{eqnarray}
\psi(v) \opsimeq_{v \to - \infty}    \frac{ -1  } {v } 
\label{fsspsideloc}
\end{eqnarray}

The finite-size correlation length $\xi_{FS}(T)$ defined by the value unity $\vert v \vert=1$ for the scaling variable $v$ diverges
with the essential singularity of Eq. \ref{xitessential}
\begin{eqnarray}
\ln \xi_{FS}(T) \opsimeq_{T \to T_c}  \frac{1}{\vert 1 -    \frac{T}{T_c} \vert }
\label{fssxi}
\end{eqnarray}

\section{ Statistics of the number $K$ of contacts }

\label{sec_contacts}

\subsection{ Statistics of the number $K$ of contacts in the delocalized phase }

In the delocalized phase, the number $K$ of contacts remains finite
in the limit $L \to +\infty$ : 
the probability distribution is simply the geometric distribution (Eq. \ref{pik}) 
\begin{eqnarray}
\pi^{deloc}_K(\infty) = \pi_0^{deloc}(\infty) \left[ 1- \pi_0^{deloc}(\infty) \right]^K
\label{piKdeloc}
\end{eqnarray}
with the essential singularity of $ \pi_0^{deloc}(\infty)$ near the transition (Eq. \ref{cumulativelogdelocfinite}).
As a consequence, in the critical region on the delocalized phase,
the probability distribution becomes the exponential distribution
\begin{eqnarray}
\pi^{deloc}_K(\infty) \simeq  e^{- \frac{A}{\theta}}  e^{-K e^{- \frac{A}{\theta}} }
\label{piKdelocexp}
\end{eqnarray}
In particular, the moments of the number of contacts diverge with the essential singularities
\begin{eqnarray}
K_p^{deloc}(L=\infty) \equiv \sum_{K=0}^{+\infty} K^p \pi^{deloc}_K(\infty) \simeq  p! e^{p \frac{A}{\theta}} 
\label{piKdelocmomentp}
\end{eqnarray}

\subsection{ Statistics of the number $K$ of contacts at the critical point }

When $\pi_0(L=\infty)=0$ and the loop distribution $P(l)$ is normalized (Eq. \ref{normapl}),
the probability of $K$ contacts in $L$ (Eq. \ref{piksum}) can be rewritten as the difference
\begin{eqnarray}
\pi_{K}(L) = {\rm Prob} \left(  \sum_{k=1}^K l_k \leq L  \right) 
-  {\rm Prob} \left(  \sum_{k=1}^{K+1} l_k \leq L \right)
\label{piksumdiff}
\end{eqnarray}
 
It is thus useful to introduce the probability distribution 
\begin{eqnarray}
{\cal S}_{K}(S_K) && = \sum_{l_1\geq 1} \sum_{l_2 \geq 1} ... \sum_{l_{K} \geq 1} P(l_1) P(l_2) ..  P(l_{K})  \delta\left(S_K-\sum_{k=1}^{K} l_k\right)
\label{sklk}
\end{eqnarray}
 of the sum 
\begin{eqnarray}
S_K \equiv \sum_{k=1}^K l_k
\label{lksum}
\end{eqnarray}
of $K$ independent loop lengths $l_k$, in order 
to rewrite Eq. \ref{piksumdiff} as
\begin{eqnarray}
\pi_{K}(L) = \int_0^L dS \left[ {\cal S}_{K}(S) - {\cal S}_{K+1}(S) \right] \simeq - \partial_K  \int_0^L dS {\cal S}_{K}(S)
\label{piksum}
\end{eqnarray}

Since the probability distribution of the loop length 
decays as (Eq. \ref{pgoogcriti})
\begin{eqnarray}
P^{criti}(l)  \oppropto_{l \to +\infty}   \frac{A}{l^{1+A} } 
\label{pgoogcritilev}
\end{eqnarray}
with $0<A<1$ (Eq. \ref{amplitude01}),
the L\'evy sum $S_K$ of Eq. \ref{lksum} does not grow extensively in $K$,
but more rapidly as $K^{\frac{1}{A}}$, and the appropriate rescaled variable
\begin{eqnarray}
\lambda \equiv A^{\frac{1}{A}} \frac{S_K}{K^{\frac{1}{A}}}
\label{lambda}
\end{eqnarray}
is distributed with the L\'evy stable law ${\cal L}_A(\lambda)$ of index $A$
defined by the Laplace transform
\begin{eqnarray}
\int_0^{+\infty} d\lambda e^{-s \lambda} {\cal L}_A(\lambda) = e^{-  s^A [-\Gamma(-A)] }
\label{levy}
\end{eqnarray}
It displays the power-law behavior for large $\lambda$
\begin{eqnarray}
 {\cal L}_A(\lambda) \opsimeq_{\lambda \to +\infty}   \frac{1}{\lambda^{1+A} } 
\label{levytail}
\end{eqnarray}
and the essential singularity near the origin (with some constant $I(A)$)
\begin{eqnarray}
 {\cal L}_A(\lambda) \oppropto_{\lambda \to 0}   \lambda^{-1- \frac{A}{2(1-A)}} \ 
 e^{-I(A) \lambda^{- \frac{A}{1-A}}}
\label{levyorigin}
\end{eqnarray}

Plugging the scaling form
\begin{eqnarray}
 {\cal S}_{K}(S) \simeq \frac{A^{\frac{1}{A}}}{K^{\frac{1}{A}}}   {\cal L}_A \left( \lambda = \frac{A^{\frac{1}{A}}}{K^{\frac{1}{A}}} S \right) 
\label{sklevy}
\end{eqnarray}
into Eq. \ref{piksum} yields
\begin{eqnarray}
\pi_{K}(L)  && \simeq - \partial_K  \int_0^L dS {\cal S}_{K}(S)
= - \partial_K  \int_0^{L A^{\frac{1}{A}}K^{-\frac{1}{A}}} d\lambda  {\cal L}_A(\lambda) 
\nonumber \\
&& \simeq L A^{\frac{1}{A}} \frac{1}{A} K^{-\frac{1}{A}-1} {\cal L}_A \left( L A^{\frac{1}{A}} K^{-\frac{1}{A}} \right) 
\label{piksumlevy}
\end{eqnarray}

The appropriate scaling variable is thus
\begin{eqnarray}
\kappa= \frac{K}{A L^A}= \frac{1}{\lambda^A} 
\label{klambdarescal}
\end{eqnarray}
and its probability distribution is obtained from the L\'evy stable law ${\cal L}_A $ as
\begin{eqnarray}
{\cal K}_A(\kappa)= \frac{1}{A \kappa^{1+\frac{1}{A}}} {\cal L}_A \left( \kappa^{-\frac{1}{A}}  \right)
\label{kofkappa}
\end{eqnarray}
It is regular near the origin  (Eq \ref{levytail})
\begin{eqnarray}
{\cal K}_A(\kappa) \opsimeq_{\kappa \to 0}  \frac{1}{A } 
\label{levytailk}
\end{eqnarray}
and displays the following decay for large $\kappa$ (Eq. \ref{levyorigin})
\begin{eqnarray}
{\cal K}_A(\kappa) \opsimeq_{\kappa \to +\infty} \frac{1}{A }   \kappa^{-1+ \frac{1}{2  (1-A)}} \ 
 e^{-I(A) \kappa^{ \frac{1}{1-A}}}
\label{levyorigink}
\end{eqnarray}

The simplest example corresponds to the value $A=\frac{1}{2}$, where the L\'evy stable law has a simple explicit expression
\begin{eqnarray}
\int_0^{+\infty} d\lambda e^{-s \lambda} {\cal L}_{\frac{1}{2}}(\lambda) && = e^{-  \sqrt{s} [2 \sqrt{\pi}] }
\nonumber \\
 {\cal L}_{\frac{1}{2}}(\lambda) && = \frac{e^{-\frac{\pi}{\lambda}}}{\lambda^{\frac{3}{2}}}
\label{levydemi}
\end{eqnarray}
so that the scaling function is simply the half-Gaussian
\begin{eqnarray}
{\cal K}_{\frac{1}{2}}(\kappa)= \frac{2}{ \kappa^3} {\cal L}_{\frac{1}{2}} \left( \kappa^{-2}  \right)
=2 e^{-\pi \kappa^2}
\label{lambdak}
\end{eqnarray}

In summary, the probability distribution of the number $K$ of contacts in $L$ follows the scaling form
\begin{eqnarray}
 \pi^{criti}_K(L) \simeq \frac{1}{A L^A} {\cal K}_A\left( \kappa=\frac{K}{A L^A} \right)
\label{pikcriti}
\end{eqnarray}
The important point is that it scales sub-extensively with respect to the length $L$ as $K \propto L^A$,
and that it remains distributed.

As a consequence, the contact density per unit length that represents the order parameter of the transition (Eq. \ref{contactdensity})
can be rewritten as
\begin{eqnarray}
k_L \equiv \frac{K}{L} = \frac{A}{L^{1-A}} \kappa
\label{contactdensitycriti}
\end{eqnarray}
where $\kappa$ is distributed with the probability distribution ${\cal K}_A(\kappa) $ discussed above.
The fact that it remains distributed over samples, even if it corresponds to a spatial average,
is in agreement with the general phenomenon of lack of self-averaging at random critical points
\cite{domany95,AH,domany} : 
outside criticality, where there exists a finite correlation length $\xi$,
the densities of extensive thermodynamic observables are self-averaging,
because a large sample can be divided into nearly independent large sub-samples of size $\xi$;
however at criticality, this 'subdivision' argument breaks down
because of the divergence of the correlation length $\xi$ \cite{domany95,AH,domany}.

\subsection{ Statistics of the number $K$ of contacts in the localized phase }

In the localized phase, the probability distribution of the number $K$ of contacts in $L$ reads (Eq. \ref{piksum})
\begin{eqnarray}
\pi^{loc}_{K}(L)  \simeq - \partial_K  \int_0^L dS {\cal S}_{K}(S)
\label{piksumloc}
\end{eqnarray}
where ${\cal S}_{K}(S_K)  $ is the probability distribution
 of the sum of $K$ independent loop lengths
\begin{eqnarray}
S_K \equiv \sum_{k=1}^K l_k
\label{lksumloc}
\end{eqnarray}
distributed with  the stretched exponential distribution of Eq. \ref{pgoodloc}
\begin{eqnarray}
P^{loc}(l)  \opsimeq_{l \to +\infty}   \frac{A}{l^{1-t} } e^{-  \frac{A l^t }{ t } }
\label{pgoodlock}
\end{eqnarray}
so that all moments are finite (Eq. \ref{nploc}).

In the typical region, the Central Limit Theorem holds 
\begin{eqnarray}
{\cal S}_{K}(S) \opsimeq_{typ}  \frac{1}{\sqrt{2 \pi K \Delta^2 }} e^{- \frac{(S-K l_{av}^{loc} )^2 }{2 K \Delta^2}}  
\label{clt}
\end{eqnarray}
in terms of the averaged loop length (Eq. \ref{navloc})
\begin{eqnarray}
l_{av}^{loc} \equiv \sum_{l=1}^{+\infty} l P^{loc}(l) 
&& \opsimeq_{t \to 0}  \sqrt{ \frac{2 \pi }{t}}  e^{ \frac{1}{t} \left[ A-1-\ln A  \right] }
\label{navlock}
\end{eqnarray}
and of the variance (Eq. \ref{nploc})
\begin{eqnarray}
\Delta^2 \equiv  \sum_{l=1}^{+\infty} l^2 P^{loc}(l)  - ( l_{av}^{loc})^2 = 
&& \opsimeq_{t \to 0}  \sqrt{ \frac{4 \pi }{t}}  e^{ \frac{2}{t} \left[ \frac{A}{2}-1-\ln \frac{A}{2}  \right] }
\label{nplocvar}
\end{eqnarray}
Plugging Eq. \ref{clt} into Eq. \ref{piksumloc} yields
\begin{eqnarray}
\pi^{loc}_{K}(L)  && \opsimeq_{typ} - \partial_K  \int_{-\infty}^L dS \frac{1}{\sqrt{2 \pi K \Delta^2 }} e^{- \frac{(S-K l_{av}^{loc} )^2 }{2 K \Delta^2}}  
\simeq - \partial_K  \int_{-\infty}^{\frac{L-K l_{av}^{loc}  }{ \sqrt{ K} \Delta}} du \frac{1}{\sqrt{2 \pi  }} e^{- \frac{u^2 }{2 }}  
\nonumber \\
&& \simeq \frac{(L+K l_{av}^{loc} )}{ 2 \Delta K^{\frac{3}{2}} \sqrt{2 \pi  }} e^{- \frac{(L-K l_{av}^{loc} )^2 }{2 K \Delta^2}}  
\label{piksumlocclt}
\end{eqnarray}

The probability distribution $G_L(k_L) $ of the contact density $k_L=\frac{K}{L} $ (Eq. \ref{contactdensity})
reads
\begin{eqnarray}
G_L(k) = L \pi^{loc}_{K=Lk}(L)  
\simeq L^{\frac{1}{2}}  \frac{(1+ k l_{av}^{loc} )}{ 2 \Delta  k^{\frac{3}{2}} \sqrt{2 \pi  }} e^{- L\frac{  ( l_{av}^{loc} )^2 }{2  k \Delta^2}\left(k-\frac{1}{ l_{av}^{loc} }\right)^2}  
\label{probcontactdensityloc}
\end{eqnarray}
In the large $L$ limit, it becomes concentrated around $k \simeq  \frac{1}{ l_{av}^{loc} }$ 
and can be approximated by the Gaussian distribution
\begin{eqnarray}
G_L(k) \opsimeq_{typ}  \sqrt{    \frac{ L ( l_{av}^{loc} )^{3} } {  2 \pi   \Delta^2 } } e^{- L\frac{  ( l_{av}^{loc} )^3 }{2   \Delta^2}\left(k-\frac{1}{ l_{av}^{loc} }\right)^2}  
\label{probcontactdensitylocgauss}
\end{eqnarray}

Besides these Gaussian fluctuations in the typical region
 around the typical value $k_{typ} =\frac{1}{ l_{av}^{loc} } $,
one may also consider the large deviation properties (see the review \cite{touchette})
and ask for the probability of an anomalously large contact density $k=\frac{K}{L}$
far from the typical value. For instance the maximal possible value $k=1$ occurs 
only if all the $L$ random contact energies of the sample turn out to be positive,
which happens with the exponentially small probability (Eq. \ref{amplitude})
\begin{eqnarray}
G_L(k=1) = \left[ \int_0^{+\infty} d \epsilon \rho(\epsilon) \right]^L = A^L = e^{-L (-\ln A) }
\label{largedevk1}
\end{eqnarray}
More generally, in the whole region $k_{typ}=\frac{1}{ l_{av}^{loc} } \leq k \leq 1$, one expects the large-deviation form 
\begin{eqnarray}
G_L( k) \oppropto_{ k_{typ}\leq k \leq 1 }  e^{- L r(k)}
\label{largedevkbigger}
\end{eqnarray}
involving a rate function $r(k)$
that interpolates between the Gaussian form of Eq. \ref{probcontactdensitylocgauss} and the limiting value $r(k=1)=-\ln A$
of Eq. \ref{largedevk1}.
Note however that the probability of an anomalously small contact density $0 \leq k<k_{typ}$ does not follow the usual large deviation form of Eq. \ref{largedevkbigger}, since we have already seen that the probability of $K=0$ contacts only decay as the stretched exponential (Eq. \ref{cumulativelogloc}).

\subsection{ Finite-size scaling for the averaged number of contacts in the critical region }

\label{sec_fsskav}

In the critical region, the above results concerning the averaged number $K_{av}(L)$ of contacts in $L$
can be summarized by the finite-size scaling form analogous to Eq. \ref{fss} involving the same correlation length of Eq. \ref{fssxi}
\begin{eqnarray}
\ln K_{av}(L) &&  \simeq   A \ln L \ \ \Psi \left[ v \equiv \left( 1-\frac{T}{T_c} \right) \ln L \right]
\label{fsskav}
\end{eqnarray}
where the scaling function satisfies $\Psi(v=0)=1$ to reproduce the critical behavior
\begin{eqnarray}
\ln K^{criti}_{av}(L)  \simeq   A \ln L 
\label{fsskavcriti}
\end{eqnarray}
decays as $\Psi(v \to -\infty) \propto \frac{1}{v} $ to reproduce to delocalized behavior
\begin{eqnarray}
\ln K^{deloc}_{av}(L)  &&  \simeq   \frac{A }{\frac{T}{T_c} -1} 
\label{fsspsidelock}
\end{eqnarray}
and behaves as $\Psi(v \to +\infty) \propto \frac{1}{A} \left[ 1- \frac{A-1-\ln A}{v}   \right]$
 to reproduce the localized behavior
\begin{eqnarray}
\ln K^{loc}_{av}(L) &&  \simeq    \ln L - \frac{A-1-\ln A}{  \left( 1-\frac{T}{T_c} \right)  }  
\label{fsspsilock}
\end{eqnarray}

\section{ Statistical properties of the free-energy }

\label{sec_free}

\subsection{ Statistics of the free-energy for a fixed number $K$ of contacts }

For a single 'good-enough' contact, the gain $f$ of Eq. \ref{x1gain}
is distributed with the exponential distribution
\begin{eqnarray}
p_1(f) = \mu e^{- \mu f} \theta(f \geq 0)
\label{Qexp}
\end{eqnarray}
of parameter
\begin{eqnarray}
\mu \equiv \frac{T}{W}
\label{mu}
\end{eqnarray}
In particular at the critical temperature $T_c$ of Eq. \ref{tc}, it depends only on the exponent $c$ of Eq. \ref{boucle}
\begin{eqnarray}
\mu_c = \frac{T_c}{W} =\frac{1}{c}
\label{mucriti}
\end{eqnarray}

The sum
\begin{eqnarray}
F_K  = \sum_{k=1}^K f_k
\label{fksum}
\end{eqnarray}
of $K$ independent gains $f_k$ can be written for any $K$ as the convolution of $K$ exponential distribution $p_1$
\begin{eqnarray}
p_K(F) = \mu \frac{(\mu F)^{K-1} }{(K-1)!} e^{- \mu F} \theta(F \geq 0)
\label{qexpK}
\end{eqnarray}

The exponential moments of order $q$ exist only in the region $0<q<\mu$ and read
\begin{eqnarray}
\overline{ e^{qF_K} } = \left( \overline{ e^{qf} } \right)^K=\left(  \int_0^{+\infty} df e^{qf} p_1(f) \right)^K
= \left( \frac{\mu}{\mu-q} \right)^K
\label{qexpKmoments}
\end{eqnarray}
The fact that each exponential moment of order $q$ involves a different exponential behavior
\begin{eqnarray}
\overline{  e^{qF_K} } = e^{ K \lambda_q }
\label{zkq}
\end{eqnarray}
with
\begin{eqnarray}
\lambda_q =\ln \frac{\mu}{\mu-q}
\label{lambdaq}
\end{eqnarray}
can be understood from the large-deviation analysis \cite{touchette} 
of the probability distribution ${\cal P}_K(g)$
of the rescaled variable $g=\frac{F_K}{K}$ for large $K$ :
using the Stirling approximation
\begin{eqnarray}
 (K-1)!   \opsimeq_{ K \to +\infty} \sqrt{2 \pi (K-1)}
 \left( \frac{K-1}{e} \right)^{K-1} 
\label{stirling}
\end{eqnarray}
one obtains
\begin{eqnarray}
{\cal P}_K(g)&& =K p_K( K g)
 \opsimeq_{K \gg 1} K \mu e^{- K \mu  g} \left[ \frac{e \mu K g}{K-1 }  \right]^{K-1}
= K \mu e^{- K \mu  g} \left[e \mu g \left(1+ \frac{1}{K-1 } \right) \right]^{K-1}
\nonumber \\
&& \opsimeq_{K \gg 1} \frac{K}{g}  \left[e \mu g e^{-\mu g} \right]^{K} 
\opsimeq_{K \gg 1} e^{K h(g) } 
\label{largedevg}
\end{eqnarray}
where the large deviation function and its two first derivatives read
\begin{eqnarray}
h(g) && = \ln \left[e \mu g e^{-\mu g} \right] = 1 + \ln (\mu g) - \mu g
\nonumber \\
h'(g) && = \frac{1}{g}-\mu
\nonumber \\
h''(g) && = -\frac{1}{g^2}
\label{largedev}
\end{eqnarray}
The expansion up to second order around the maximum
\begin{eqnarray}
g_0=\frac{1}{\mu}
\label{largedevgtyp}
\end{eqnarray}
where it vanishes $h(g_0)=0$ correspond to the Gaussian distribution
of the Central-Limit theorem in the typical region.
The whole large-deviation function is however necessary to evaluate
the exponential moments of arbitrary order $0<q<\mu$ 
\begin{eqnarray}
 \overline{ e^{q F_K} }
\oppropto_{K \gg 1}  \int dg e^{ K \left[ h(g)+ q g \right] }
\label{momentsqzk}
\end{eqnarray}
The integral is dominated by the saddle-point value $g_q$ satisfying
\begin{eqnarray}
0 = h'(g_q)+q =  \frac{1}{g}-\mu+q
\label{saddlegq}
\end{eqnarray}
yielding
\begin{eqnarray}
g_q=\frac{1}{\mu-q}
\label{largedevgq}
\end{eqnarray}
and one indeed recovers the exponential behavior of Eq. \ref{zkq} with
\begin{eqnarray}
\lambda_q = h(g_q)+ q g_q =   1 + \ln (\mu g_q) - \mu g_q+q g_q =\ln \frac{\mu}{\mu-q}
\label{lambdaqbis}
\end{eqnarray}
This shows that the moments of order $0<q<\mu$ are dominated by the atypical values $q_q$ of Eq. \ref{largedevgq}
bigger than the typical value $g_0$.

\subsection{ Statistics of the free-energy for a fixed length  $L$ }

The moments of the partition function (Eq \ref{momentq}) of arbitrary order $0<q<\mu$ read using Eq. \ref{qexpK}
\begin{eqnarray}
M_q(L) \equiv \overline{Z_L^q} && 
= \sum_{K=0}^{L} \pi_K(L)   \left( \frac{\mu}{\mu-q} \right)^K
\label{zlq}
\end{eqnarray}

The series expansion in $q$ allows to compute the moments of $F_L\equiv \ln Z_L$ (Eq. \ref{freeenergy})
\begin{eqnarray}
\sum_{p=0}^{+\infty}
\frac{q^p}{p!} \overline{ F_L^p  }  &&
= \sum_{K=0}^L  \pi_K(L) \left(1- \frac{q}{\mu} \right)^{-K} =  \sum_{K=0}^L \pi_K(L)
\sum_{p=0}^{+\infty}\left( \frac{q}{\mu} \right)^p \frac{(K+p-1)! }{p! (K-1)!}
\label{zlqseries}
\end{eqnarray}
The identification order by order yields that the integer moment of order $p$
\begin{eqnarray}
\overline{ F_L^p  }   && = \frac{1}{\mu^p}  \sum_{K=0}^L  \pi_K(L)
 \frac{(K+p-1)! }{ (K-1)!} = \frac{1}{\mu^p}  \sum_{K=0}^L \pi_K(L) \prod_{j=0}^{p-1} (K+j)
\label{freemomentp}
\end{eqnarray}
involves the moments up to order $p$ of the number of contacts
\begin{eqnarray}
K_p(L)= \sum_{K=0}^L \pi_K(L) K^p
\label{kpmoment}
\end{eqnarray}

In particular, the averaged free-energy is directly related to the averaged number of contacts
$K_{av}(L)=K_1(L) $
\begin{eqnarray}
\overline{ F_L  } =  \frac{ K_{av}(L) }{\mu} 
\label{logzav}
\end{eqnarray}
and thus inherits the finite-size scaling properties discussed in Eq. \ref{fsskav}.

The variance of the free-energy
\begin{eqnarray}
\overline{ F_L^2  }  - \left( \overline{ F_L  }  \right)^2 
= \frac{K_{av}(L)+ K_2(L)-K_{av}^2(L) }{\mu^2}
\label{logz2}
\end{eqnarray}
is also directly related to the average $K_{av}(L) $ and variance $[K_2(L)-K_{av}^2(L) ]$ of the number of contacts.

\section{ Statistics of the free-energy in the delocalized phase }

\label{sec_freedeloc}

\subsection{Probability distribution of the free-energy in the limit $L \to +\infty$} 

In the delocalized phase, the number $K$ of contacts remains finite
in the limit $L \to +\infty$ (Eq. \ref{piKdeloc}).
As a consequence, the partition function $Z_L$ 
and the free-energy $F(L)=\ln Z(L)$ remain also finite random variable as $L \to +\infty$.

The probability distribution of $F_{L}=\ln Z_{L}$ in the limit $L \to +\infty$ reads
\begin{eqnarray}
{\cal F}^{deloc}_{L=\infty}(F) && = \sum_{K=0}^{\infty} \pi_K^{deloc}(\infty) p_K(F) 
\nonumber \\
&& = \pi_0^{deloc}(\infty) \delta(F) +\sum_{K=1}^{\infty}\pi_0^{deloc}(\infty)\left[ 1- \pi_0^{deloc}(\infty) \right]^K
 \mu \frac{(\mu F)^{K-1} }{(K-1)!} e^{- \mu F} \theta(F \geq 0)
 \nonumber \\
&& = \pi_0^{deloc}(\infty) \delta(F) +\left[ 1- \pi_0^{deloc}(\infty) \right] \mu  \pi_0^{deloc}(\infty) 
 e^{ - \pi_0^{deloc}(\infty)  \mu F } \theta(F \geq 0) 
\label{freeprobadeloccal}
\end{eqnarray}

Near the transition where $  \pi_0^{deloc}(L=\infty)$  is given by Eq \ref{cumulativelogdelocfinite} in terms of the reduced temperature $\theta \equiv \frac{T}{T_c}-1 >0 $,
the probability distribution of the free-energy $F \geq 0$ of Eq. \ref{freeprobadeloccal} becomes the exponential distribution 
\begin{eqnarray}
{\cal F}^{deloc}_{L=\infty}(F) && \simeq \alpha(\theta)
 e^{ -  \alpha(\theta) F } 
\label{freeprobadeloc}
\end{eqnarray}
of parameter
\begin{eqnarray}
\alpha(\theta) \equiv \mu  \pi_0^{deloc}(L=\infty) \opsimeq_{\theta \to 0} \mu_c e^{- \frac{A}{\theta}}
\label{mutheta}
\end{eqnarray}
that vanishes with an essential singularity near the transition.

For the partition function $Z=e^{F}$, the exponential distribution of Eq. \ref{freeprobadeloc}
translates into the power-law distribution for the partition function $Z \geq 1$
\begin{eqnarray}
{\cal Z}^{deloc}_{L=\infty}(Z) && \simeq \frac{\alpha(\theta)}{Z^{1+  \alpha(\theta) } }
\label{zprobadeloc}
\end{eqnarray}
with a L\'evy exponent $\alpha(\theta)$ that becomes very small near the transition (Eq. \ref{mutheta}).

\subsection{ Moments of the free-energy $F_L=\ln Z_L$ in the limit $L \to +\infty$} 

The moments of the free-energy distributed with the exponential distribution of Eq. \ref{freeprobadeloc} are simply
\begin{eqnarray}
\overline{F_{L=\infty}^p}=   \frac{ p! }{ \left[ \alpha(\theta) \right]^p} 
\label{fpdeloc}
\end{eqnarray}
In particular, the averaged value diverges with the essential singularity (Eq. \ref{mutheta})
\begin{eqnarray}
\overline{F_{L=\infty}}= \overline{(\ln Z_{\infty})} 
&& =  \frac{ 1 }{ \alpha(\theta) } \oppropto_{\theta \to 0} \frac{1}{\mu_c} e^{\frac{A}{\theta} } 
\label{freeavdeloc}
\end{eqnarray}
as the variance
\begin{eqnarray}
\overline{F_{\infty}^2} - (\overline{F_{\infty}} )^2 = \frac{ 1  }{ \left[ \alpha(\theta) \right]^2} 
\oppropto_{\theta \to 0} \frac{1}{\mu_c^2} e^{2\frac{A}{\theta} } 
\label{freevardelocdeloc}
\end{eqnarray}

\subsection{ Moments of the partition function} 

The moments of the partition function distributed with Eq. \ref{zprobadeloc}
\begin{eqnarray}
M^{deloc}_q(L=\infty) && = \frac{ 1 }{ 1 -\frac{q}{ \alpha(\theta) }  }
\label{zlqdeloc}
\end{eqnarray}
exist only in the region
\begin{eqnarray}
0< q< \alpha(\theta) \propto \mu_c e^{- \frac{A}{\theta}}
\label{qcdeloc}
\end{eqnarray}
that is shrinking to zero at the critical point is approached.

\section{Statistics of the free-energy at the critical point }

\label{sec_freecriti}

\subsection{ Moments of the free-energy $F_L=\ln Z_L$ } 

At criticality, the moments of the number of contacts
\begin{eqnarray}
K^{criti}_p(L) && =  \int dK K^p   \pi^{criti}_K(L) 
\nonumber \\
&& \opsimeq_{L \to +\infty}   (A L^A)^p \int_0^{+\infty} dk k^p {\cal K}_A\left( k \right)
\label{kplcriti}
\end{eqnarray}
can be plugged into Eq. \ref{freemomentp} to obtain that the moments of the free-energy scale as
\begin{eqnarray}
\overline{ F_L^p  }   &&  = \frac{1}{\mu_c^p}  \sum_{K=0}^L \pi_K(L) \prod_{j=0}^{p-1} (K+j)
\nonumber \\
&& \opsimeq_{L \to +\infty}    \left( \frac{A L^A }{\mu_c} \right)^p \int_0^{+\infty} dk k^p {\cal K}_A\left( k \right)
\label{freemomentpcriti}
\end{eqnarray}

In particular, one obtains that the averaged value
\begin{eqnarray}
\overline{ F_L  } =  \frac{ K_{av}(L) }{\mu_c} =   L^A \ \frac{ A }{\mu_c} \int_0^{+\infty} dk k {\cal K}_A\left( k \right)
\label{logzavcriti}
\end{eqnarray}
and the width
\begin{eqnarray}
\sqrt{ \overline{ F_L^2  }  - \left( \overline{ F_L  }  \right)^2 }
\opsimeq_{L \to +\infty}  L^A \ \frac{ A }{\mu_c}  \sqrt{\left[  \int_0^{+\infty} dk k^2 {\cal K}_A\left( k \right) 
- \left( \int_0^{+\infty} dk k {\cal K}_A\left( k \right)  \right)^2  \right] }
\label{logz2criti}
\end{eqnarray}
both scale as $L^A$.

The conclusion is thus that the free-energy density can be rewritten as
\begin{eqnarray}
f_L \equiv \frac{F_L}{L} = \frac{A}{ \mu_c L^{1-A}  } \kappa
\label{freedensitycriti}
\end{eqnarray}
where $\kappa$ remains distributed over samples with the probability distribution ${\cal K}_A(\kappa) $ of Eq. \ref{kofkappa}. Again, the fact that it remains distributed over samples
is in agreement with the general lack of self-averaging at random critical points
\cite{domany95,AH,domany} mentioned after Eq. \ref{contactdensitycriti}.

For the partition function $Z_L=e^{F_L}$, this corresponds to the typical scaling behavior
\begin{eqnarray}
Z_L \opsimeq_{typ} e^{ \frac{A}{ \mu_c  } L^A \kappa }
\label{ztypcriti}
\end{eqnarray}
whereas the moments display a completely different scaling as we now discuss.

\subsection{ Moments of the partition function $Z_L$ } 

The scaling distribution (Eq. \ref{pikcriti}) 
\begin{eqnarray}
 \pi^{criti}_K(L) \simeq \frac{1}{A L^A} {\cal K}_A\left( k=\frac{K}{A L^A} \right)
\label{plzcritibis}
\end{eqnarray}
can be plugged into Eq. \ref{zlq} for the moments of order $q<\mu_c$ to obtain
\begin{eqnarray}
M^{criti}_q(L) \equiv \overline{Z_L^q } &&
= \int dK \pi_K(L)   \left( \frac{\mu_c}{\mu_c-q} \right)^K
\simeq \int_0^{+\infty} d\kappa  {\cal K}_A(\kappa) e^{  A L^A \kappa \ln \left( \frac{\mu_c}{\mu_c-q} \right)}
\label{zlqcriti}
\end{eqnarray}
The divergence for large $L$ is thus governed by the asymptotic behavior of 
Eq. \ref{levyorigink} for large $\kappa$.
The change of variable
\begin{eqnarray}
 \kappa=L^{1-A} v
\label{kappavanomalous}
\end{eqnarray}
leads to
\begin{eqnarray}
M^{criti}_q(L) 
\simeq \int_0^{+\infty} d\kappa  \kappa^{-1+ \frac{1}{2  (1-A)}}    \ 
 e^{-I(A) \kappa^{ \frac{1}{1-A}}}
 e^{  A L^A \kappa \ln \left( \frac{\mu_c}{\mu_c-q} \right)}
\nonumber \\
\simeq L^{\frac{1}{2}} \int_0^{+\infty} dv    v^{ -1 + \frac{1}{2  (1-A)}} \ 
 e^{ L \left[ -I(A) v^{ \frac{1}{1-A}}+  A v \ln \left( \frac{\mu_c}{\mu_c-q} \right)
\right] }
\label{mlqcriti}
\end{eqnarray}
The saddle-point analysis thus involves the function
\begin{eqnarray}
\phi(v) && = -I(A) v^{ \frac{1}{1-A}}+  A v \ln \left( \frac{\mu_c}{\mu_c-q} \right)
\nonumber \\
\phi'(v) && = -I(A)\frac{1 }{1-A} v^{ \frac{A}{1-A}}+  A  \ln \left( \frac{\mu_c}{\mu_c-q} \right)
\nonumber \\
\phi''(v) && = -I(A)\frac{A }{(1-A)^2} v^{ \frac{2A-1}{1-A}}
\label{phikappa}
\end{eqnarray}

The saddle-point value $v_q$ where $\phi'(v_q)=0 $ reads
\begin{eqnarray}
 v_q = \left[  \frac{A (1-A) }{I(A)} \ln \left( \frac{\mu_c}{\mu_c-q} \right)\right]^{\frac{1-A}{A}}
\label{vq}
\end{eqnarray}
and leads to the values
\begin{eqnarray}
\phi(v_q) && = -I(A) v_q^{ \frac{1}{1-A}}+  A v_q \ln \left( \frac{\mu_c}{\mu_c-q} \right)
= \frac{A  I(A) } { 1-A } 
\left[  \frac{A (1-A) }{I(A)} \ln \left( \frac{\mu_c}{\mu_c-q} \right)\right]^{\frac{1}{A}}
\label{phikappaq}
\end{eqnarray}
and
\begin{eqnarray}
\phi''(v_q) && = -I(A)\frac{A }{(1-A)^2} v_q^{ \frac{2A-1}{1-A}}
=  -I(A)\frac{A }{(1-A)^2}
\left[  \frac{A (1-A) }{I(A)} \ln \left( \frac{\mu_c}{\mu_c-q} \right)\right]^{\frac{2A-1}{A}}
\label{phikappaqderi2}
\end{eqnarray}

So the saddle-point evaluation of the integral of Eq. \ref{mlqcriti} 
\begin{eqnarray}
M^{criti}_q(L) && \opsimeq_{L \to +\infty}  L^{\frac{1}{2}}
 \sqrt{ \frac{2 \pi}{- L \phi''(v_q)}}   v_q^{ \frac{2A-1}{2  (1-A)}} \   e^{L \phi(v_q) }
 \oppropto_{L \to +\infty}   e^{L \phi(v_q) }
\label{zlqcritires}
\end{eqnarray}
corresponds to an exponential divergence in $L$ completely different from the scaling of the typical free-energy of Eq. \ref{logzavcriti}.
The difference can be traced back to the fact that the typical samples correspond to finite values of the variable $\kappa$,
i.e. to a number of contacts scaling as $K \propto L^A\kappa$,
while the saddle-point evaluation in the variable $v$ of Eq. \ref{kappavanomalous}
is dominated by the rare sample having an anomalously large $\kappa \propto L^{1-A} v$
corresponding to an extensive number of contacts $K \propto L^A\kappa \propto L v$.

As a consequence, the limit $q \to 0$ is singular for the above saddle-point computation
\begin{eqnarray}
 v_q  \opsimeq_{q \to 0} \left[  \frac{A (1-A) }{I(A)} \frac{q}{\mu_c} \left(1+ \frac{q}{2\mu_c}+O(q)) \right)\right]^{\frac{1-A}{A}}  \oppropto_{q \to 0} q^{\frac{1-A}{A}}
\nonumber \\
\phi(v_q) \opsimeq_{q \to 0} \frac{A  I(A) } { 1-A } 
 \left[  \frac{A (1-A) }{I(A)} \frac{q}{\mu_c} \left(1+ \frac{q}{2\mu_c}+O(q)) \right)\right]^{\frac{1}{A}}
   \oppropto_{q \to 0} q^{\frac{1}{A}}
\label{kappaq}
\end{eqnarray}
and the series expansion in $q$ to recover the typical behavior of the partition function (Eq. \ref{ztypcriti})  is not possible.

It is interesting to interpret the anomalous behavior of the moments of Eq. \ref{zlqcritires}
in terms of large deviations \cite{touchette} :
 at criticality, even if the typical free-energy scales sub-extensively as $F_{typ} \propto L^A$,
the probability to have a finite free-energy density $f=\frac{F}{L}$ is exponentially small in $L$
and follows the large-deviation form with some rate function $\psi(f)$
\begin{eqnarray}
Prob( F=L f) \propto e^{-L \psi(f)}
\label{largedevfree}
\end{eqnarray}
Then the moments of the partition function corresponds to the saddle-point evaluation
\begin{eqnarray}
M^{criti}_q(L) &&  \simeq \int df e^{L [- \psi(f)+q f]}  \propto e^{L [- \psi(f_q)+q f_q]}
\label{zlqcritiresld}
\end{eqnarray}
where the saddle-point value $f_q$ corresponds to
\begin{eqnarray}
\psi'(f_q)=q
\label{largedevfq}
\end{eqnarray}
so that the rate function $\psi(f)$ corresponds to the Legendre transform of 
\begin{eqnarray}
\lambda(q) = - \psi(f_q)+q f_q = \phi(v_q) 
&& = \frac{A  I(A) } { 1-A } 
\left[  \frac{A (1-A) }{I(A)} \ln \left( \frac{\mu_c}{\mu_c-q} \right)\right]^{\frac{1}{A}}
\label{phivq}
\end{eqnarray}

\section{Statistics of the free-energy in the localized phase }

\label{sec_freeloc}

\subsection{ Statistics of the free-energy density $f_L=\frac{F_L}{L}$ } 

In the localized phase, 
the averaged free-energy grows extensively as the averaged number of contacts
\begin{eqnarray}
\overline{ F_L  } =  \frac{ K^{loc}_{av}(L) }{\mu} \opsimeq \frac{ L }{\mu \   l^{loc}_{av} } 
\label{logzavloc}
\end{eqnarray}
and the variance grows also extensively as 
\begin{eqnarray}
\overline{ F_L^2  }  - \left( \overline{ F_L  }  \right)^2 = \frac{K_{av}(L)+ K_2(L)-K_av^2(L) }{\mu^2} 
\opsimeq_{L \to +\infty}   \frac{L}{\mu^2 l^{loc}_{av}  } \left[  1+ \frac{\Delta^2}{( l^{loc}_{av} )^2} \right]
\label{logz2loc}
\end{eqnarray}

In the typical region, the probability distribution of the free-energy density $f=\frac{F_L}{L}$ 
thus follows the Gaussian distribution
\begin{eqnarray}
{\cal P}^{loc}_L(f)  \opsimeq_{typ}  \sqrt{    \frac{ L l_{av}^{loc} \mu^2 } {  2 \pi  \left[  1+ \frac{\Delta^2}{( l^{loc}_{av} )^2} \right]  } } 
e^{- L\frac{   l_{av}^{loc}  \mu^2 }{2  \left[  1+ \frac{\Delta^2}{( l^{loc}_{av} )^2} \right] }\left(f  -\frac{1}{ \mu \  l_{av}^{loc} }\right)^2}  
\label{probfreedensitylocgauss}
\end{eqnarray}

For the partition function $Z_L=e^{F_L}$, this translates into the log-normal distribution
in the typical region
\begin{eqnarray}
P_L^{loc}(Z) \opsimeq_{typ} \frac{1}{Z}  \sqrt{    \frac{  l_{av}^{loc} \mu^2 } {  2 \pi  \left[  1+ \frac{\Delta^2}{( l^{loc}_{av} )^2} \right] L } } 
e^{- \frac{   l_{av}^{loc}  \mu^2 }{2  \left[  1+ \frac{\Delta^2}{( l^{loc}_{av} )^2} \right]L }\left( \ln Z  -\frac{L}{ \mu \  l_{av}^{loc} }\right)^2}  
\label{ztyploc}
\end{eqnarray}
while the moments will be governed by the large deviation sector as we now discuss.

\subsection{ Moments of the partition function $Z_L$ }

The moment of the partition function (Eq. \ref{zlq}) of order $q>0$
can be rewritten as an integral over the contact density $0\leq k = \frac{K}{L} \leq 1$
\begin{eqnarray}
M_q(L) \equiv \overline{Z_L^q} && 
= \sum_{K=0}^{L} \pi_K(L)   \left( \frac{\mu}{\mu-q} \right)^K
\nonumber \\
&& = L \int_0^{1}  dk \pi_{kL}(L) e^{L k \ln \left( \frac{\mu}{\mu-q} \right) }
\label{zlqloc}
\end{eqnarray}
This integral will thus involve the large deviation form of Eq. \ref{largedevkbigger}
\begin{eqnarray}
G_L( k) =L \pi_{kL}(L) \oppropto_{ k_{typ}\leq k \leq 1 }  e^{- L r(k)}
\label{largedevkbiggerbis}
\end{eqnarray}
describing the exponentially small probability to have a bigger contact density $k$
than the typical one $k_{typ}$.
The saddle-point evaluation
\begin{eqnarray}
M_q(L) \propto \int_0^{1}  dk  e^{L \left[ k \ln \left( \frac{\mu}{\mu-q} \right) -r(k) \right] } \propto e^{L \left[ k_q \ln \left( \frac{\mu}{\mu-q} \right) -r(k_q) \right] } 
\label{zlqlocsad}
\end{eqnarray}
is dominated by the saddle point value $k_q$ such that
\begin{eqnarray}
r'(k_q) = \ln \left( \frac{\mu}{\mu-q} \right) 
\label{kqsaddle}
\end{eqnarray}

\section{ Validity of the Strong Disorder Renewal Approach }

\label{sec_validity}

\subsection{ Notion of Strong Disorder Fixed Point }

From the point of view of Strong Disorder Approaches \cite{review_igloi},
random critical points can be decomposed into : 

(i)  ``Infinite Disorder Fixed Points'', as introduced by Daniel Fisher 
\cite{danielrtfic,danielantiferro,danielreview},
where Strong Disorder Approaches become asymptotically exact because the
 disorder width becomes larger and larger with the scale and thus dominate over quantum, thermal, or stochastic fluctuations\cite{review_igloi};

(ii) ``Finite Disorder Fixed Points'', where the disorder width remains finite at large scale.
Then Strong Disorder Approaches are not asymptotically exact but are expected to become good approximations
in the region where the disorder width is sufficiently large (see various examples in the two reviews
\cite{review_igloi,review_refael} and references therein) : the best known example is the Griffiths phases around 
``Infinite Disorder Fixed Points'' analyzed by Daniel Fisher \cite{danielrtfic,danielantiferro,danielreview},
or equivalently the anomalous diffusion phase $x \propto t^{\mu}$ with $0<\mu<1$ of the biased Sinai model 
where the results obtained by Strong Disorder Renormalization \cite{rg_sinai,rg_trap} 
can be compared to results obtained by other methods (see the reviews \cite{review_georges} and references therein ).

For our present wetting or DNA denaturation model, we have found that the free-energy gain $f$ of a 
single 'good-enough' contact is distributed at criticality with the exponential distribution of Eq. \ref{Qexp} 
with the parameter $\mu_c=\frac{1}{c} $ of Eq. \ref{mucriti} : this corresponds to the case (ii) of ``Finite Disorder Fixed Points'',
and the Strong Disorder Approximation is expected to become better for small $\mu_c$, i.e. for large exponent $c$,
corresponding to large dimension $d$ for the case of random walks (Eq. \ref{cRW}).
The experience with the biased Sinai model mentioned above indicates that the
Strong Disorder Approach could actually give the correct critical behaviors in the whole region $0<\mu_c<1$
that would correspond to any loop exponent $c>1$.

\subsection{ Correspondence with the quantum long-ranged Ising chain with random transverse fields }

There exists some partial correspondence between the present classical wetting model and the one-dimensional
random transverse field long-ranged Ising model 
 as studied by the Strong Disorder Renormalization procedure in Ref \cite{igloi_lr}.
The exponent $\alpha$ governing the power-law decay of the couplings $J_{ij} \propto \vert i-j \vert^{-\alpha}$ of Ref \cite{igloi_lr}
corresponds to the exponent $c$ of the loop weight $\Omega(l) = l^{-c}$
(Eq \ref{boucle}) of the wetting model 
\begin{eqnarray}
\alpha=c
\label{alphac}
\end{eqnarray}
The random transverse fields $h_i$ of Ref \cite{igloi_lr} corresponds to the contact energies of the wetting model via 
\begin{eqnarray}
\frac{\epsilon_i }{T}= - \ln h_i
\label{fieldhi}
\end{eqnarray}
 Although the correspondence between the two models is not complete (in particular the quantum spins 
are subjected to all the transverse fields $h_i$ and to all the couplings $J_{ij}$ between pairs, while the classical polymer makes loops),
the relation becomes explicit when both models are analyzed via Strong Disorder Renormalization.
Indeed, the Strong Disorder RG rules within the 'primary model' described in Ref \cite{igloi_lr} 
for the random transverse field long-ranged Ising model 
can be translated for the wetting model as follows :

(i)  The elementary contributions to the free energy are the contacts free-energies $f_{n_i}=\frac{\epsilon_{n_i} }{T}= - \ln h_{n_i} $
and the loop entropic costs $f_{n_in_{i+1}}=- \ln \Omega(n_{i+1}-n_i)=-c \ln (n_{i+1}-n_i) = - \ln J_{n_in_{i+1}}$.

(ii) The decimation of the strongest parameter $(h_{n_i},J_{n_jn_{j+1}})$ in the quantum model 
corresponds to the decimation of the smallest free-energy contribution $(f_{n_i},f_{n_jn_{j+1}})$.

(iii) When the site of the smallest parameter $f_{n_{i_0}}$ is decimated, this bad contact and its two neighboring loops
are replaced by a single loop characterized by free-energy contribution $f_{n_{i_0-1},n_{i_0+1}}= -c \ln (n_{i_0+1}-n_{i_0-1}) $.

(iii) When the loop of the smallest parameter $f_{n_{i_0},n_{i_0+1}}$ is decimated, this loop and its two neighboring contacts
are replaced by a single contact of free-energy contribution $f^{new}= f_{n_{i_0}}-c \ln (n_{i_0+1}-n_{i_0}) +f_{n_{i_0+1}} $.

 This correspondence at the level of Strong Disorder RG rules explains why the same essential singularity of the correlation length (Eq. \ref{xitessential}) appears,
and why the dynamical exponent $z_c=\alpha$ governing the power-law behavior of the renormalized random fields $g(h) \propto h^{\frac{1}{z_c}-1}$ for $h \to 0$ at the critical point of the quantum chain \cite{igloi_lr}
corresponds to our notation $\mu_c=\frac{1}{c}=\frac{1}{z_c}$ in Eq. \ref{Qexp}.

As a final remark concerning the validity of Strong Disorder Approaches,
it is interesting to mention that the authors of Ref \cite{igloi_lr} have studied numerically the three values $\alpha=2,3,4$
 that correspond to the values $c=2,3,4$ for the loop exponent of the wetting model.

\section{ Conclusion}

\label{sec_conclusion}

For the random DNA denaturation transition, or equivalently the random wetting transition, we have introduced a Strong Disorder Renewal Approach to construct the optimal contacts in each disordered sample. We have analyzed the statistics of the loop lengths, of the number of contacts and of the free-energy over the ensemble of disordered samples of a given length $L$. The correlation length governing the finite-size scaling properties in the critical region has been found to diverge with the essential singularity of Eq. \ref{xitessential} discussed in the Introduction. At the critical point, we have found that both the contact density (order parameter) and the free-energy density decay as a power-law of the length $L$ but remain distributed, in agreement with the general phenomenon of lack of self-averaging at random critical points \cite{domany95,AH,domany}. We have obtained that for any real $q>0$, the moment $\overline{Z_L^q} $ of order $q$ of the partition function $Z_L$ is dominated at criticality by some exponentially rare samples displaying a finite free-energy density, i.e. by the large deviation sector of the probability distribution of the free-energy density.

Further work is needed to understand the origin of the difference with the BKT scenario of Eq. \ref{xiBKT}
found by the real-space renormalization procedures on hierarchical lattices in Refs \cite{tang,retaux}.

\section*{ Acknowledgments}

It is a pleasure to thank Bernard Derrida and Martin Retaux 
for sending me the PhD Thesis based on their joint work \cite{retaux} that has rekindled my interest 
in the random wetting transition after many years far from this field.

\appendix

\section{ Application to other distributions of the contact energies } 

\label{sec_app}

In the main text, we have focused on the exponential distribution of Eq. \ref{rhoexp}
for the contact energies, because it simplifies the technical details. However, it is important to explain in this Appendix
how the Strong Disorder Renewal Approach can be adapted to other distributions $\rho(\epsilon)$.
The main idea is that even if the distribution $\rho(\epsilon)$ of an { \it individual } contact energy $\epsilon$
decays more rapidly than exponentially, an exponential tail will be nevertheless generated in the probability
distribution of the contact energy of {\it good segments } \cite{tang},  as is also well known
in the context of Strong Disorder Renormalization Approaches \cite{review_igloi}.
In the following, we explain how this phenomenon occurs within the particular details of the wetting model that we consider.

\subsection{ Initial coarse-graining  }

Let us perform an initial coarse graining as follows :
the consecutive sites having a positive contact energy $\epsilon >0$ are grouped together into 'attractive' segments,
while the consecutive sites having a negative contact energy $\epsilon <0$ are grouped together into 'repulsive' segments.
In terms of the probabilities of positive and negative contact energies
\begin{eqnarray}
p \equiv \int_0^{+\infty} d\epsilon \rho(\epsilon) 
\nonumber \\
1-p \equiv \int_{-\infty}^0 d\epsilon \rho(\epsilon) 
\label{petp}
\end{eqnarray}
the probability distributions $X(x)$ and $Y(y)$ of the lengths $x=1,2,..$ and $y=1,2,..$ of attractive and repulsive segments
reads respectively
\begin{eqnarray}
X(x) && =(1-p) p^{x-1}
\nonumber \\
Y(y) && =p (1-p)^{y-1}
\label{xybinary}
\end{eqnarray}
Then a disorder realization $(\epsilon_1,\epsilon_2,...)$ of individual contact energies 
 is recast into as a series of segments of lengths $(y_1,x_1,y_2,x_2,...)$ draw with Eq. \ref{xybinary}.

For a repulsive segment, the only important variable is the length $y$.
But an attractive segment is characterized by both its length $x$ and its contact energy
\begin{eqnarray}
E = \sum_{j=1}^x \epsilon_j
\label{rhosegment}
\end{eqnarray}
where the positive $\epsilon_j$ are drawn with the probability distribution
\begin{eqnarray}
\rho_{+}(\epsilon) \equiv  \frac{\rho(\epsilon) \theta(\epsilon \geq 0) }{p} 
\label{rhoplus}
\end{eqnarray}

The probability distribution ${\cal X}(E)$ of the energy $E$ of an attractive segment of any length $x$ reads
\begin{eqnarray}
{\cal X}(E)= \sum_{x=1}^{+\infty} X(x)  \int_0^{+\infty} d\epsilon_1 \rho_+(\epsilon_1) ... \int_0^{+\infty} d\epsilon_x \rho_+(\epsilon_x) \delta( E - \sum_{j=1}^x \epsilon_j)
\label{rhosegmentjoint}
\end{eqnarray}
It is thus convenient to work with the Laplace transforms
\begin{eqnarray}
{\hat \rho_+}(\lambda )
\equiv \int_0^{+\infty} d\epsilon \rho_+(\epsilon) e^{-\lambda \epsilon} = \frac{1}{p}  \int_0^{+\infty} d\epsilon \rho(\epsilon) e^{-\lambda \epsilon} 
\label{rhohat}
\end{eqnarray}
and
\begin{eqnarray}
{\hat {\cal X}} (\lambda)  && \equiv \int_0^{+\infty} dE e^{-\lambda E} {\cal X}(E)
\label{xhat}
\end{eqnarray}
to translate Eq. \ref{rhosegmentjoint}
into
\begin{eqnarray}
{\hat {\cal X}} (\lambda)  && =  (1-p)  \sum_{x=1}^{+\infty} p^{x-1} \left[ {\hat \rho_+}(\lambda ) \right]^x  
=  \frac{ (1-p) {\hat \rho_+}(\lambda ) }{1-p {\hat \rho_+}(\lambda ) }
\label{xhatlap}
\end{eqnarray}

In this Appendix, we consider that $\rho(\epsilon)$ decays more rapidly than exponentially, so that $ {\hat \rho_+}(\lambda )$
exists for any $\lambda \in ]-\infty,+\infty[$. However Eq. \ref{xhatlap} displays a pole at the negative value $\lambda_c= - \frac{1}{W}$ satisfying
\begin{eqnarray}
1= p {\hat \rho_+}\left(\lambda_c= - \frac{1}{W} \right) = \int_0^{+\infty} d\epsilon \rho(\epsilon) e^{ \frac{ \epsilon} {W}}
\label{pole}
\end{eqnarray}

The residue of the pole at $\lambda_c$ in Eq. \ref{xhatlap}
reads
\begin{eqnarray}
{\rm Res}({\hat {\cal X}} (\lambda); \lambda_c)  && =   \frac{ (1-p) {\hat \rho_+}(\lambda_c ) }{-p {\hat \rho_+}'(\lambda_c ) }
=  \frac{ (1-p) \frac{1}{p} }{  \int_0^{+\infty} d\epsilon \rho(\epsilon) \epsilon e^{ \frac{\epsilon} {W}}}
\label{residu}
\end{eqnarray}
The Tauberian theorem then yields that ${\cal X}(E) $ decays as
\begin{eqnarray}
{\cal X}(E) \opsimeq_{E \to +\infty}  {\rm Res}({\hat {\cal X}} (\lambda); \lambda_c)  e^{\lambda_c E}
\label{tauberian}
\end{eqnarray}

In summary, the probability to have an energy $E$ bigger than some large threshold $\eta$ displays the exponential decay 
(analogous to Eq. \ref{rhotail})
\begin{eqnarray}
\int_{\eta}^{+\infty} dE {\cal X}(E) \opsimeq_{\eta \to +\infty}  B e^{- \frac{\eta}{W}}
\label{rhosegmentlarge}
\end{eqnarray}
where the parameter $W$ and the amplitude $B$ are computed from the initial distribution $\rho(\epsilon)$ by 
\begin{eqnarray}
  \int_0^{+\infty} d\epsilon \rho(\epsilon) e^{ \frac{ \epsilon} {W}} =1
\label{poleW}
\end{eqnarray}
and 
\begin{eqnarray}
B= W \ {\rm Res}({\hat {\cal X}} (\lambda); \lambda_c)  && =   \frac{ (1-p) W }{ p \int_0^{+\infty} d\epsilon \rho(\epsilon) \epsilon e^{ \frac{\epsilon} {W}}}
\label{residuamplib}
\end{eqnarray}

\subsection{ Strategy in each disordered sample }

The Strong Disorder strategy described in section \ref{sec_strategy} can be now adapted as follows.
The first segment $y_1$ with negative contact energy is 'bad' by definition.
The second segment $x_1$ is considered as 'good enough' if
its contact energy $E_1$ is greater than the entropic cost due to the previous bad segment $y_1$
\begin{eqnarray}
E_1 > c T \ln (y_1)
\label{firstb}
\end{eqnarray}
If this condition is not satisfied, one asks whether $x_2$ is a 'good enough segment'
 satisfying
\begin{eqnarray}
E_2 > c T \ln (y_1+x_1+y_2)
\label{secondb}
\end{eqnarray}
and so on. So the first good-enough segment $x_{n_1^*}$ 
corresponds to the first $n$ satisfying
\begin{eqnarray}
E_n > c T \ln \left( \sum_{i=1}^n y_i+\sum_{i=1}^{n-1} x_i \right)
\label{secondbn}
\end{eqnarray}
and the corresponding gain for the logarithm of the partition function is
\begin{eqnarray}
f_{n_1^*}= \frac{E_{n_1^*}}{T} -c \ln \left( \sum_{i=1}^{n_1^*} y_i+\sum_{i=1}^{{n_1^*}-1} x_i \right) \geq 0
\label{x1gainb}
\end{eqnarray}

\subsection{ Probability $\pi_0(L=\infty)$ of zero contact on the half-infinite line }

The probability $\pi_0(L=\infty)$ to find zero 'good enough segment' on the half-infinite line satisfies (as Eq. \ref{cumulativelog})
\begin{eqnarray}
\ln \pi_0(L=\infty) = - \sum_{n=1}^{+\infty} u(n)
\label{cumulativeb}
\end{eqnarray}
where the elementary term $u(n)$ replacing Eq. \ref{un} reads
\begin{eqnarray}
u(n) && = - \ln \left[ 1- Prob \left( E_n >c T \ln \left( \sum_{i=1}^n y_i+\sum_{i=1}^{n-1} x_i \right) \right) \right]
\nonumber \\
&& = - \ln \left[ 1-  \prod_{i=1}^n (\sum_{y_i=1}^{+\infty} Y(y_i))  \prod_{j=1}^{n-1} ( \sum_{x_j=1}^{+\infty} X(x_j) ) 
\int_{c T \ln \left( \sum_{i=1}^n y_i+\sum_{i=1}^{n-1} x_i \right)}^{+\infty} dE {\cal \chi}(E) ] \right]
\label{unbb}
\end{eqnarray}

For large $n$, it is convenient to introduce the rescaled variable
\begin{eqnarray}
r  \equiv \frac{1}{n} \left( \sum_{i=1}^n y_i+\sum_{k=1}^{n-1} x_i \right) 
\label{rsumlargen}
\end{eqnarray}
and its probability distribution $R_n(r)$ that becomes concentrated around the averaged value
\begin{eqnarray}
\overline{r} = \overline{y_i}+\overline{ x_i}  = \frac{1}{p} +  \frac{1}{1-p}   =\frac{1}{p (1-p)}
\label{sumlargen}
\end{eqnarray}

Using the asymptotic behavior of Eq. \ref{rhosegmentlarge}
 the asymptotic behavior of Eq. \ref{unbb} is given by
\begin{eqnarray}
u(n)  &&= - \ln \left[ 1-  \int dr R_n(r) \int_{c T \ln \left( nr \right)}^{+\infty} dE {\cal \chi}(E) ] \right]
\nonumber \\
&& \opsimeq_{n \to +\infty} \int dr R_n(r) B e^{- \frac{c T }{W} \ln \left( nr \right) } =  \frac{B \int dr R_n(r)  r^{- \frac{c T }{W}  } }{ n^{\frac{cT}{W} } }
\nonumber \\
&& \opsimeq_{n \to +\infty}    \frac{B [p(1-p) ]^{ \frac{c T }{W}  } }{ n^{\frac{cT}{W} } }
\label{unbblarge}
\end{eqnarray}
This corresponds to a power-law form analog to Eq. \ref{ularge}
\begin{eqnarray}
u(n) \opsimeq_{n \to +\infty}  \frac{ A(T) }{n^{ \frac{ T }{T_c} } }
\label{ulargeb}
\end{eqnarray}
with the critical temperature $T_c=\frac{W}{c}$ as in Eq. \ref{tc}
and the amplitude that now depends on the temperature
\begin{eqnarray}
A(T)= B \left[ p (1-p) \right]^{\frac{T}{T_c} } 
\label{amplitudetemp}
\end{eqnarray}

\subsection{ Probability $p_1(f)$ of the free-energy gain $f$ of a 'good-enough segment' }

From Eq. \ref{rhosegmentlarge}, one obtains that the probability distribution of the free-energy gain of Eq. \ref{x1gainb}
 decays exponentially as
\begin{eqnarray}
p_1(f) \oppropto_{f \to +\infty}  \mu e^{- \mu f}
\label{segmentgaindecay}
\end{eqnarray}
with the parameter $\mu=\frac{T}{W}$ as in Eq. \ref{mu}.
At the critical temperature $T_c=\frac{W}{c}$ of Eq. \ref{tc}, one recovers again $ \mu_c = \frac{T_c}{W} =\frac{1}{c}$ as in Eq. \ref{mucriti}.

\subsection{ Special case of the binary distribution }

The case of the binary distribution,
where  the contact energies can take only two values $\pm \epsilon_0$ with probabilities $p$ and $(1-p)$ respectively
\begin{eqnarray}
\rho(\epsilon) = p \delta( \epsilon - \epsilon_0 ) + (1-p) \delta( \epsilon +\epsilon_0 ) 
\label{rhobinary}
\end{eqnarray}
 is more natural in the context of DNA denaturation where there are two types of base pairs.
This case
has to be treated slightly differently, because the energy of a good segment of length $x$ is now exactly 
proportional to its length $E=x \epsilon_0$ and is thus a discrete variable distributed with
\begin{eqnarray}
{\cal X}(E)= \sum_{x=1}^{+\infty} X(x)  \delta( E - x\epsilon_0) = (1-p) \sum_{x=1}^{+\infty} p^{x-1}  \delta( E - x\epsilon_0) 
\label{rhosegmentjointbinary}
\end{eqnarray}
So the probability to have an energy $E$ bigger than some large threshold $\eta$ 
corresponds to the probability to have a length $x \geq {\rm Int}(\frac{\eta}{\epsilon_0})+1 $
\begin{eqnarray}
\int_{\eta}^{+\infty} dE {\cal X}(E) && =  (1-p) \sum_{x={\rm Int}(\frac{\eta}{\epsilon_0})+1 }^{+\infty} p^{x-1}  = p^{{\rm Int}(\frac{\eta}{\epsilon_0})}
\opsimeq_{\eta \to +\infty} e^{- \frac{\eta}{\epsilon_0} \ln \frac{1}{p} }
\label{rhosegmentlargebinary}
\end{eqnarray}
so that the parameters $W$ and $B$ of Eq \ref{rhosegmentlarge} read
\begin{eqnarray}
W = \frac{ \epsilon_0}{\ln \frac{1}{p} }
\label{Wbinary}
\end{eqnarray}
and 
\begin{eqnarray}
B=   1
\label{bbinary}
\end{eqnarray}

\subsection{ Conclusion of the Appendix}

The conclusion of this Appendix is that the Strong Disorder Renewal Approach described in the main text
for the simple case of an exponential distribution of the individual contact energies, can be adapted to other probability distributions that decay
more rapidly than exponentially, provided one performs an initial coarse graining into attractive and repulsive segments :
an exponential tail of rare good segments is then generated with the parameters derived above,
 and the analysis of the main text can be performed mutatis mutandis, without changing the critical scalings.

\end{document}